# Sen2Chain: An Open-Source Toolbox for Processing Sentinel-2 Satellite Images and Producing Time-Series of Spectral Indices

Authors: Christophe Révillion[1,2], Pascal Mouquet[1,2], Jérémy Commins[1,2,3], Juliette Miranville[1,2], Charlotte Wolff[1,2], Thomas Germain[1,2], Sylvaine Jégo[1,2], Lucas Longour[1,4], Florian Girond[5], Didier Bouche[6], Rodolphe Devillers[1,2], Gwenaëlle Pennober[1,2], Vincent Herbreteau[1,2,4]

[1] Espace-Dev, IRD, Univ Montpellier, Univ Guyane, Univ La Réunion, Univ Antilles, Univ Nouvelle Calédonie, Montpellier France

[2] Espace-Dev, Univ La Réunion, Saint Denis, La Réunion, France

[3] Office national des forêts, Direction départementale de Guyane, Cayenne, Guyane, France

[4] Khmer Earth Observation Laboratory, Institute of Technology of Cambodia, Phnom Penh, Cambodia

[5] Communicable Disease Control Department, Ministry of Health, Phnom Penh, Cambodia

[6] DSI, Université de La Réunion, Station SEAS-OI, Saint-Pierre, La Réunion, France

Corresponding author: Christophe Révillion, christophe.revillion@univ-reunion.fr, Université de La Réunion 40 avenue de Soweto, 97410 Saint-Pierre, La Réunion

Link to the code: https://framagit.org/espace-dev/sen2chain

## Authorship contribution statement

All authors have read and agree to the submitted version of the manuscript. Conceptualization C.R., P.M. and V.H.; Funding acquisition G.P. and V.H.; Methodology C.R., C.W., J.C., J.M., P.M. and V.H.; software development C.W., J.C., J.M., P.M., and S.J.; tool testing and operational use C.R., D.B., F.G, L.L., P.M., T.G. and V.H.; software documentation C.R., J.C., P.M. and T.G., Writing - original draft C.R and P.M; Writing. - review & editing C.R., C.W., D.B., F.G., G.P., J.C., J.M., L.L., P.M., R.D., S.J., T.G. and V.H.

## Abstract

The increasing availability of free high-resolution earth observation data covering any point on the globe every few days led to the emergence of new remote sensing tools that can manipulate the very large volumes of data generated by those satellites. We present 'Sen2Chain', an open-source Python tool that can automate the processing of large time series of Sentinel-2 images for their use in various fields (e.g., environmental health, natural hazards, ecology). Sen2Chain allows downloading images from various earth observation data suppliers, applying geometric and atmospheric corrections using ESA's 'Sen2Cor' tool, and generating and applying cloud masks. Sen2Chain's ability to extract time series of spectral indices (e.g NDVI, NDWI) provides simplified access to value-added environmental information for a wide range of end-users and applications.

Sen2Chain enables all data processing stages to be customized and chained together, with the possibility to automate and parallelize the processing, and optimize data management. Sen2Chain is paving the way for the creation and processing of a large earth observation image database dedicated to users who require time-series and/or perform regular environmental observations. The Web tool 'Sen2Extract' is also presented, which enables end-users with no expertise in remote sensing to easily extract time-series for 11 spectral indices values for specific regions of interest.

# 1 Introduction

Since the first images of the Earth were acquired from space, Earth Observation (EO) by remote sensing has become a powerful tool for understanding natural and anthropogenic processes. EO data are constantly improving, with the launch of satellites and technological advances providing higher spatial, spectral, and temporal resolutions (higher revisit frequency). These data are becoming increasingly accessible through user-friendly data catalogs, Application Programming Interfaces (API), and free and often open data tools (Vancauwenberghe and van Loenen, 2018). This has been the case since the United States Geological Survey (USGS) began distributing Landsat optical images for free in 2008 and more recently, in 2014, with the launch of the European Space Agency's (ESA) Copernicus program and its constellation of Sentinel satellites. This free access to massive data has paved the way for increased operational uses in many fields, such as environmental monitoring, ecology, agriculture, disaster management, and health. However, using such extensive EO datasets requires suitable tools and processing chains that can exploit them efficiently in an automated way (Salamon et al., 2021; Roberts et al., 2022), especially to enable the generation of time-series of standardized and calibrated data.

Characterizing individual pixel values over time using various spectral indices is common in remote sensing research (Hislop et al., 2018). Generating time-series of environmental indices from EO data (e.g., vegetation, water or humidity indices) is essential to monitor and predict land surface variations. It is widely applied in land cover change detection and monitoring ecosystem dynamics. These time-series can capture seasonal environmental fluctuations and local dynamics that may be linked to human practices such as deforestation (Gao et al., 2020), agriculture (Bégué et al., 2018), water quality (Pahlevan et al., 2022), fire monitoring and mapping (Huang et al., 2016; Verhegghen et al., 2016), but also exceptional weather events such as floods or cyclones (Sanyal and Lu, 2004; Mouquet et al., 2020). EO data characteristics vary based on the type of sensor used (i.e. optical, radar), which can be influenced by acquisition conditions such as cloud cover and atmospheric composition (Girard and Girard, 2010). Fine spatial resolution time-series can also be used to complement low-resolution data from meteorological satellites (on the order of a kilometer) but whose acquisition frequency is much higher (i.e., daily to hourly) (Ceccato et al., 2005; Marais Sicre et al., 2016).

The capacity to produce times series of EO data using the free Sentinel-2 (S2) optical satellite constellation (ESA's Copernicus program) has marked a clear milestone, bringing a significant gain in both spatial (down to 10 meters) and temporal (one image every 5 days) resolutions, when compared to previous satellites. These characteristics have already proven their worth in a wide range of fields, such as the mapping of land-cover (Phiri et al., 2020), the impact of natural disasters (Mouquet et al., 2020), and the mapping of habitats (Poursanidis et al., 2019).

Sentinel-2 data users are provided with advanced levels of data preprocessing (L1C images are already orthorectified for geometric sensor and relief effects) and free access to the ESA atmospheric calibration tool 'Sen2Cor'. This, combined with the ease of access to S2 data (direct online access via Web portals or API), opened prospects for scientific research and the development of operational products.

Such a context encouraged further processing of entire time series of images, including developing a new software tool that can automate the generation of environmental indices from these images. This article presents Sen2Chain, a software tool developed as a processing chain to download, manage, and process large volumes of S2 data covering broad temporal and spatial extents.

## 2 Existing tools for accessing and processing S2 data

Following the launch of the two S2 satellites in 2015 and 2017, the EO community has been heavily involved in facilitating access and processing of these data. This section provides an inventory of existing tools and identifies gaps that can inform the development of new functionalities. An extensive list of tools and portals dedicated to S2 data is also available on the "Awesome-sentinel GitHub page".

### 2.1 Data access

ESA has set up the Copernicus Data Space, a portal part of the Copernicus program, giving free and open access to data from the entire Sentinel satellite constellation. This portal provides access to images from all the Sentinel constellation satellites (i.e., Sentinel-1, Sentinel-2, Sentinel-3, Sentinel-5, and Sentinel-6). S2 data can be accessed in raw (L1C) or pre-processed (L2A) formats applying atmospheric corrections, as generated by the Sen2Cor tool (Louis et al., 2016).

Several other services offer access to all or part of the S2 database, with access rights, availability, and update frequency varying amongst operators. This includes the '*Plateforme d'Exploitation des Produits Sentinel*' (PEPS) of the Centre National d'Etudes Spatiales (CNES), where all data are made available for free at the L1C processing level. Other mirror portals offer access to all or part of the S2 catalog. Many of these access nodes can be configured using the open-source Earth Observation Data Access Gateway (EODAG) tool developed by CSGroup that lets you search for and download S2 and other data from a list of providers.

### 2.2 Data preprocessing

Copernicus supplies S2 data in the L1C processing level that were corrected for the geometric effects of sensor distortion, projection, and relief (Main-Knorn et al., 2017). These orthorectified images can then be overlaid onto each other and on a map. The next step in processing the data series is to correct for atmospheric effects to obtain Top Of Canopy (TOC) data. This results in L2A level data that are radiometrically comparable and hence suitable for temporal comparisons. Several tools are available for this step, including MAJA developed by CNES and CESBIO (Hagolle et al., 2015) and Sen2Cor developed by ESA (Louis et al., 2016). Both tools are free and open-source, the main difference being in the correction algorithm used. MAJA is based on a multi-temporal pre-processing method, enabling the outputs to be refined using several images over the same area. This makes the tool more efficient locally, notably for detecting clouds and shadows, but also implies a more complex implementation, as a minimum number of images over the same area is required to run the pre-processing efficiently. Sen2Cor can be used for cloud detection but can be less robust when detecting diffuse clouds such as cirrus clouds, or highly reflecting objects such as urban areas or beach tops (Baetens et al., 2019). Sen2Cor is regularly updated and its algorithms improved, particularly for cloud detection.

2.3 More comprehensive tools for processing Sentinel-2 data

Two general approaches are found to process large time series of EO data: cloud-based platforms such as Google Earth Engine (GEE) (Gorelick et al., 2017) and open-source systems. GEE is a powerful and increasingly popular EO online platform that provides access to large ready-for-use EO datasets, to a wide range of tools for manipulating them and can be used to develop and run codes addressing specific user needs. GEE provides access to the large storage and computing power of Google' servers and is free of use for research and education. Accessibility is facilitated by a straightforward online Web interface, which does not require high connection speeds, making it more usable when users have limited network access, storage capacities, and computing power (Liu et al., 2020; Rosa et al., 2023). If such a solution offers many benefits, it also comes with limitations like a limited free storage space, restrictions on the processing tools available for free, limitations in terms of sharing of data with other uses, and privacy rules governed by Google (Padarian et al., 2015; Amani et al., 2020). Future changes in the policies governing GEE could also be problematic for projects that have invested in the platform for large developments if their needs cannot be fulfilled anymore.

Open-source environments that allow handling large amounts of EO data include the 'Sen2r' package (Ranghetti et al., 2020) developed in the popular R software (R Core Team, 2018). Sen2r offers S2 image download functionalities, image pre-processing using Sen2Cor, and the ability to generate spectral indices. The tool was designed to process high-frequency time series of large raster files on "standard" desktop PCs or more powerful server infrastructures. The 'Sen2-Agri' tool (European Space Agency, 2018) written in Bash (processing tools), HTML, and JavaScript (interface) allows generating data mainly useful in agronomic applications (e.g., monthly composite images, NDVI and LAI maps, monthly crop type maps). Sen2Agri presents limits to customize products and requires specific hardware to be run (Ranghetti et al., 2020).

## 3. General architecture of the Sen2Chain tool

### 3.1 Sen2Chain overview

Sen2Chain is a software tool developed as a processing chain to download, manage, and process large volumes of S2 data covering broad temporal and spatial extents. It was designed to make it easy for anyone, regardless of programming skills, to manage a large amount of EO data, to automate the processing of S2 images, and generate time series of radiometric indices. The development of Sen2Chain was initiated in 2017 during research projects that dealt with cyclone impact mapping (Tulet et al., 2021; Mouquet et al., 2020) and the use of environmental data in epidemiological surveillance systems.

Sen2Chain was entirely developed in Python, based on free software libraries developed by the user community. Python and R are two of the most widely used open-source languages in geospatial (Canty, 2014; Kaya et al., 2019). Since the 2000s, many popular free high-performance libraries have been developed in Python for raster and vector GIS data management, image processing, and remote sensing. The most common and widely used libraries include GDAL, Rasterio, Pandas, Numpy, SciPy, and MatPlotLib. Specific packages and tools were also used for handling Sentinel data, such as EODAG for downloading data and Sen2Cor for performing atmospheric corrections on images. An exhaustive list of the Python packages used and required to run Sen2Chain is provided on the tool's Gitlab page.

Sen2Chain has been designed to handle the large volume of data involved in processing such high spatial resolution high-frequency data, which can easily reach tens of thousands of images - or tens of terabytes - stored locally on the user's machine.

The basic processing unit handled by Sen2Chain is the S2 tiles that are 110x110 km$^2$ ortho-images in UTM/WGS84 projection, following the Military Grid Reference System divisions. Each of these tiles is identified by a unique identifier to which all data are associated to, from L1C images to spectral indices.

Sen2Chain allows downloading S2 L1C images from different online catalogs using the EODAG tool (Figure 1). After querying the online catalogs, downloads are performed by geographic zones, using the identifiers of the tiles of interest, and a user-defined maximum cloud cover and temporal extent. L1C images are then pre-processed, e.g. corrected for atmospheric effects and attenuations, using ESA's Sen2Cor tool to produce L2A images. Once this calibration stage has been completed, the L2A images produced, are used to generate cloud masks and spectral indices (e.g., brightness, vegetation, and humidity indices).

All processing steps in the chain have been designed and optimized to handle large volumes of data, in particular through parallelization of processes on computer cores when possible, distributing the load and reducing the computation time. For example, L1C image downloads are done in parallel depending on the possibilities offered by each provider (8 concurrent threads for PEPS, four concurrent threads for Copernicus Data Space).

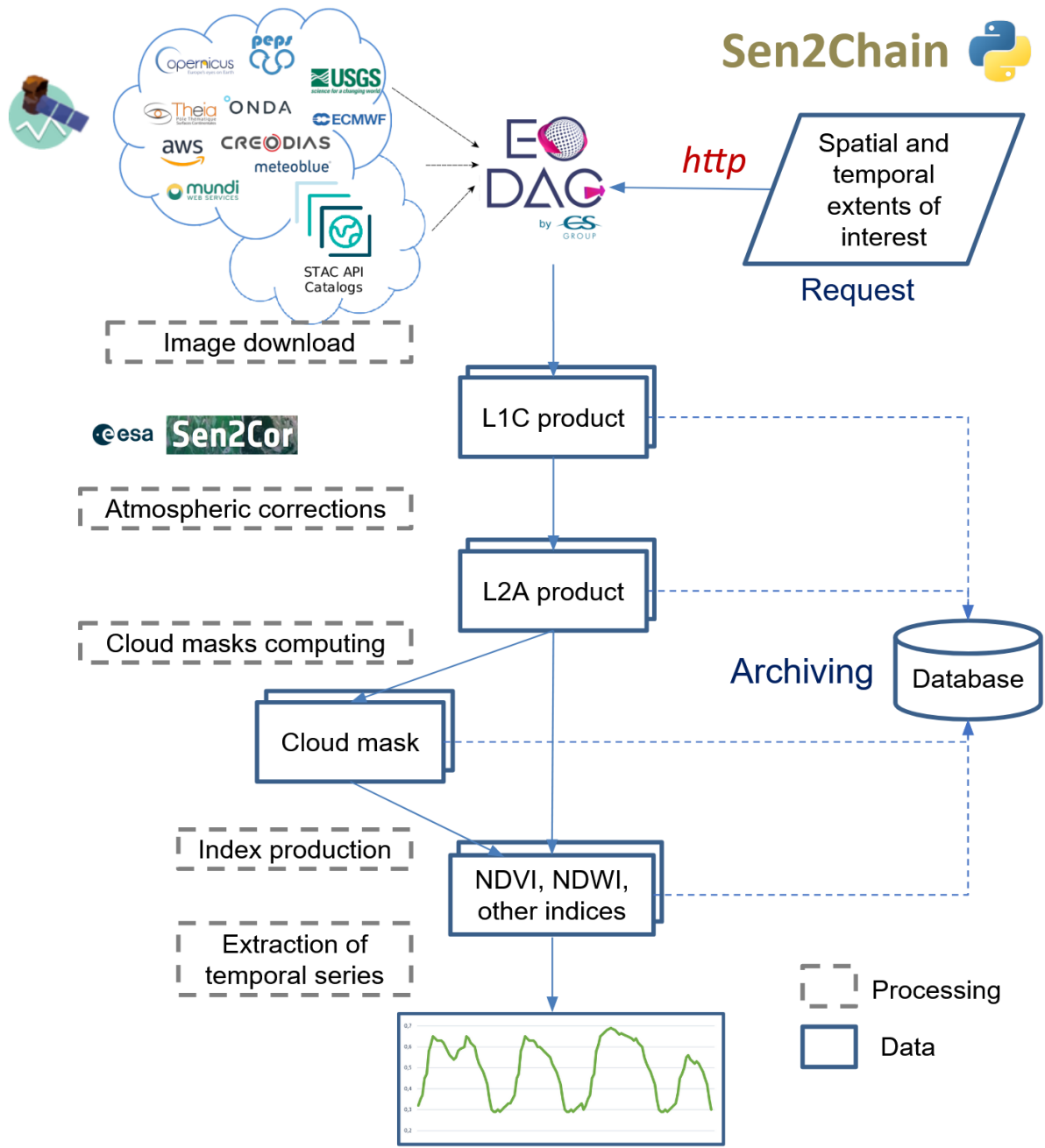

Figure 1: Main stages in downloading, processing and storing Sentinel-2 images and their products with the Sen2Chain

A specific 'Job' class has been implemented to facilitate the management of the image database and the execution of all the different processing tasks. Jobs enable a set of tiles to be processed as a whole (from downloading to index production) in a recurring and automatic manner, for a specific use or area. This feature allows the processing of large volumes of data and reduces processing times.

### 3.2 Sen2Chain key steps

#### 3.2.1 Data download

Sen2Chain uses third-party open-source software bricks to retrieve S2 images from various online data providers. Historically, Sen2Chain has relied on the Python tools 'peps_download' and 'Sentinelsat', allowing S2 data to be directly queried and downloaded from the CNES's PEPS platform and the ESA's Copernicus Open Access Hub, respectively. More recently, as protocols and providers changed and these older tools were not maintained, Sen2Chain used the EODAG tool, a well-documented open-source code developed by CS GROUP with a Python API. EODAG is actively maintained and gives access to larger data catalogs and suppliers.

As with all processing with Sen2Chain, the downloading of L1C Sentinel-2 images is done at the spatial scale of the S2 tile (110x110km). The 'Tile' class allows to perform all query operations on online databases, with the possibility of filtering images matching temporal or cloud cover criteria.

#### 3.2.2 Application of atmospheric corrections with the Sen2Cor tool

L1C images downloaded from EODAG need to be corrected for the atmospheric effects and the satellite's acquisition conditions. After comparing various atmospheric correction tools available, we chose ESA's 'Sen2Cor' for its relative ease of use and its cloud detection efficiency. Sen2Cor calibrates the data to make them comparable over space and time, by performing atmospheric and optional terrain and cirrus cloud corrections, from Top-Of-Atmosphere Level 1C input data to Bottom-Of-Atmosphere L2A reflectance images. Additionally, other layers of important information relating to 'Aerosol optical thickness', 'Water vapor', 'Scene classification map', 'Quality indicators' for cloud and 'Snow probabilities' are produced. The correction model can be parameterized to consider the characteristics of the study area, the proportion of the various atmospheric gases, and the relief (Main-Knorn et al., 2017). The Sen2Chain user can set the parameters of Sen2Cor according to the study area desired, processing objectives, and output resolutions.

As Sen2Cor is partly integrated into Sen2Chain, a sound installation and configuration are necessary for the proper functioning of the complete processing chain. Due to ongoing updates to S2 data and their associated metadata, it is important to always keep the version of Sen2Cor up to date so that the most recent data can continue to be processed.

### 3.2.3 Cloud detection and generation of masks

Optical satellite images generally include clouds and their projected shadows on the ground that impact the ability to detect phenomena on the ground. This leads to spatial and temporal gaps in satellite Earth observation data and may cause biases in images and their derivative products. Detecting clouds and their shadows on satellite images is a crucial operation in remote sensing, as it often impacts the quality of the results obtained from the analyses of those images. Depending on their characteristics, clouds are more or less opaque to light radiation, masking the target of interest or altering the signal when they are located between the sensor and the ground. Numerous detection methods and algorithms exist (Tarrio et al., 2020) that help detect pixels altered by the presence of clouds and, if necessary, to mask them.

Sen2Chain offers the possibility of choosing between four different cloud masks (cm001-004), depending on user needs (Figure 2). They are generated using two different output data layers produced by Sen2Cor, at a 20 meters spatial resolution.

Sen2Chain's default historical cloud mask is cm001. This mask uses the cloud probability (CLDPRB) layer generated by Sen2Cor (MSK_CLDPRB) during the atmospheric correction process. The MSK_CLDPRB layer estimates the probability (0-100%) that a given pixel is obstructed by clouds or optically thick aerosols (e.g., ice or snow) (Mueller-Wilm et al., 2016). The cloud mask is computed in two steps. First, a binary image is generated by applying a threshold on MSK_CLDPRB, recognizing pixels with values above as being clouds. Then, an opening - a sequence of two successive mathematical morphology operations - is performed on the thresholded data: a binary erosion followed by a binary dilation. These steps make it possible to remove small areas detected as being clouds (e.g., isolated pixels), and extend the detection area by several pixels to hide any clouds that are missed or have diffuse contours. Sen2Chain's second cloud mask, cm002, is derived from cm001 and has been developed to reduce poor cloud detection due to abnormally high cloud probability values in Sen2Cor's MSK_CLDPRB layer. These high values can occur, for example, on particularly reflective rivers or beach edges (e.g., specular surface reflections, high turbidity, presence of white sand). To eliminate these artifacts, we use S2 mid-infrared (SWIR) B11 spectral band centered on 1600 nm, which has a low sensitivity to these disturbances. The third cloud mask cm003 is based on the same sen2Cor cloud probability layer (MSK_CLDPRB at 20 m spatial resolution) and constructed according to the same scheme as cm001, but with two new parameters allowing to change the cloud probability threshold value and the number of iterations of the binary dilation operation (but unlike cm001, without prior erosion).

Finally, the last computable cloud mask - cm004 - differs from the three others by using another Sen2Cor output layer, the Scene Classification (SC) band at 20m spatial resolution. This data layer computed during the atmospheric correction process by the Sen2Cor algorithm, has four classes of interest to better detect clouds and their shadows: medium probability of cloud (CLOUD_MEDIUM_PROBABILITY), high probability of cloud (CLOUD_HIGH_PROBABILITY), cirrus cloud (THIN_CIRRUS)), and a specific class for cloud shadows (CLOUD_SHADOWS). Users can fine-tune the creation of the mask according to their needs, using or not each of these four cloud-related classes. Several iterations of a binary dilation operation can also be used to generate a final mask with greater extension to minimize the risk of taking into account altered pixels.

Sen2Chain's ability to calculate different cloud masks allows users to adapt the processing to specific characteristics of the study area or topics under investigation.

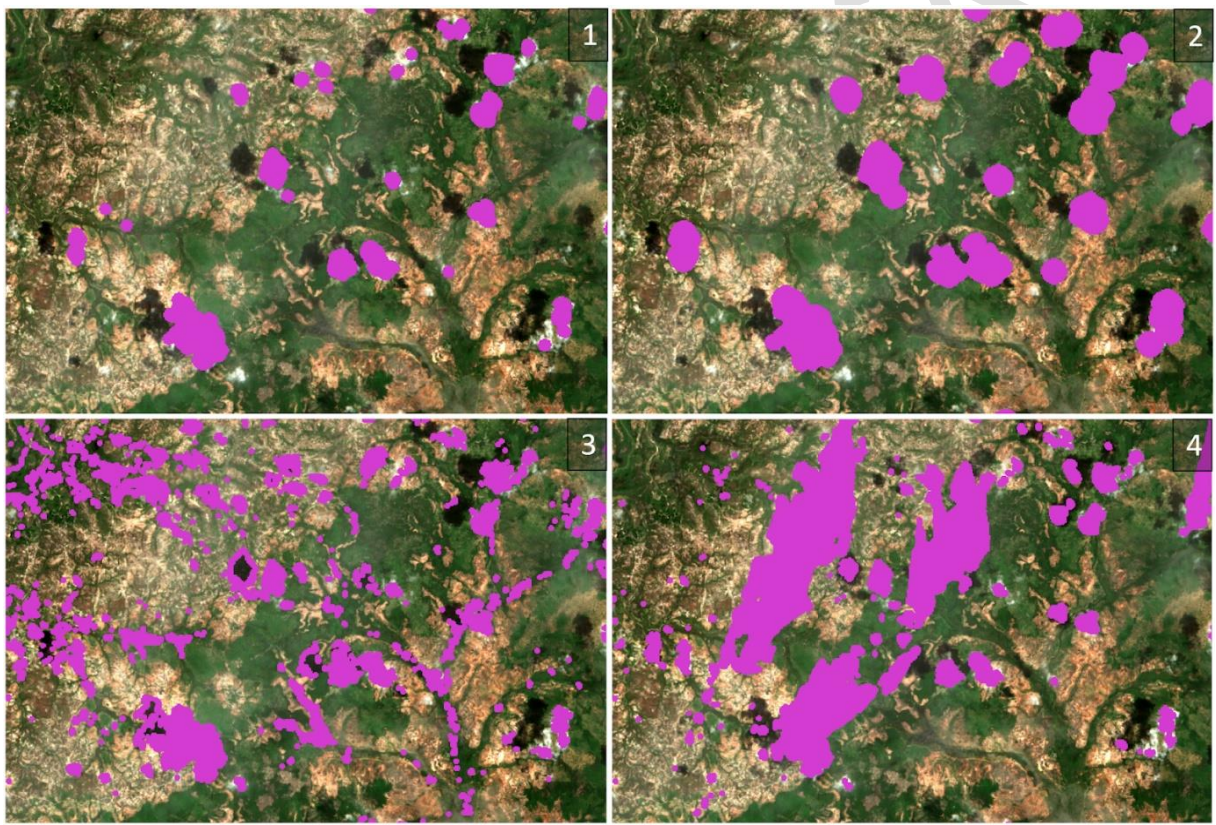

Figure 2: Example of a single Sentinel-2 image processed using the four cloud masks available in Sen2Chain: 1 - cm001; 2 - cm002; 3 - cm003 (masked zones corresponding to a cloud probability greater than 1% and one iteration of the binary dilation kernel); 4 - cm004 (using all four classes, including cirrus, and one iteration of a binary dilation kernel). Pixels detected as being clouds or their shadows are in pink (remaining white and black pixels are clouds or shadows - respectively - that have not been identified in the masks).

### 3.2.4 Radiometric indices calculation

Eleven radiometric indices can be calculated in Sen2Chain using mathematical operations on the different bands of Sentinel-2 imagery (Table 1). These include vegetation indices (NDVI, NDRE, IRECI, and EVI), moisture and water indices (NDWIGAO, NDWIMCF, and MNDWI), brightness indices (BIBG, BIGR, and BIRNIR and a fire index/burnt area (NBR)). The index values are multiplied by 10,000 in integer format, which considerably reduces the size of the raster files stored. All these indices can be easily generated on demand from L2A data, and masked using either of the available cloud masks. Output results are raster images at 10m spatial resolution in JPEG2000 format. Importantly, this list of indices is not static and the flexibility of Sen2Chain's open Python code allows for easy modification and addition of new indices based on user needs and applications.

Table 1: Radiometric indices computed by Sen2Chain ([*]NIR = Near Infrared; [**]SWIR = Short Wave Infrared).

| Acronym | Name | Equation | References |
|---|---|---|---|
| BIBG | Brightness Index (Blue-Green) | $bibg = \sqrt{\dfrac{(green^2 + blue^2)}{2}}$ | Crist et al. (1986) <br> Mathieu et al., (1998) |
| BIGR | Brightness Index (Green-Red) | $bibg = \sqrt{\dfrac{(green^2 + red^2)}{2}}$ | Crist et al. (1986) <br> Mathieu et al., (1998) |
| BIRNIR | Brightness Index (red-NIR[*]) | $bibg = \sqrt{\dfrac{(NIR^2 + red^2)}{2}}$ | Crist et al. (1986) <br> Mathieu et al., (1998) |
| EVI | Enhanced Vegetation Index | $2.5 \dfrac{(NIR - red)}{((NIR + 6red - 7.5blue) + 1)}$ | Huete et al. (1997) |
| IRECI | Inverted Red-Edge Chlorophyll Index | $\dfrac{(NIR - red)}{(Red\ edge\ 1/Red\ edge\ 2)}$ | Frampton et al. (2013) |
| MNDWI | Modification of Normalized Difference Water Index | $\dfrac{(green - SWIR)}{(green + SWIR)}$ | Xu (2006) |
| NBR | Normalized Burn Ratio | $\dfrac{(NIR - SWIR2)}{(NIR + SWIR2)}$ | Key and Benson (2006) |

| NDRE | Normalized Difference Red-Edge | $\frac{(NIR - rededge)}{(NIR + rededge)}$ | Barnes et al. (2000) |
| NDVI | Normalized Difference Vegetation Index | $\frac{(NIR - red)}{(NIR + red)}$ | Tucker (1979) |
| NDWIGAO | Normalized Difference Water Index - Gao method | $\frac{(NIR - SWIR)}{(NIR + SWIR)}$ | Gao (1996) |
| NDWIMCF | Normalized Difference Water Index - McFeeters method | $\frac{(green - NIR)}{(green + NIR)}$ | McFeeters (1996) |

3.2.5 Extraction of time series from vector point or polygon files

One of the goals in developing Sen2Chain was to facilitate access to information derived from remote sensing data. In addition to simplifying the generation of products, we have therefore added the ability to extract general spatial statistics (e.g., minimum, maximum, mean, standard deviation, percentiles values) on all spectral indices for specific locations (points) or regions of interest (polygons) and for specific dates or time periods. Such extractions generate text files (in CSV format) that can be easily integrated into a GIS, spreadsheet, or even R or Python languages, for further analyses.

## 4 Using Sen2Chain

Sen2Chain source code is open and available online on the Framagit software forge mirrored on the IRD forge. The installation procedure requires standard libraries, in particular for downloading and processing images, and is detailed in the online documentation.

Sen2Chain uses the object-oriented capabilities of the Python programming language (object-oriented programming - OOP) to bundle properties and behaviors into individual classes. Most functionalities, such as downloading and preprocessing images or producing indices are thus easily accessible and executable. The online documentation details the different use cases.

### 4.1 Class 'Tile'

The Tile class was implemented to perform all the necessary operations on one entire time series of images of the S2 library, at the spatial scale of one S2 tile (110 x 110 km) characterized by its unique identifier. This class incorporates different basic Python classes enabling all the products to be processed individually (L1C and L2A images, cloud masks, indices). By automating and looping the different methods, an entire dataset can be processed at once.

The main goal of this class is to help generate and maintain a local product database from online products available from different providers. The 'Tile' class thus makes it possible to execute the following tasks: i) query the online providers about the available products and download them if necessary (*get_l1c*), ii) correct L1C images for atmospheric and relief effects and to produce L2A images (*compute_l2a*), iii) generate cloud masks (*compute_cloudmasks*), and iv) generate spectral indices (*compute_indices*).

Many of the functions used in this class are based on existing open-source libraries, allowing specific actions to be carried out on S2 products. We can notably cite EODAG which allows intelligent downloading from different online data providers. The 'Tile' class also uses ESA Sen2Cor tool to correct the atmospheric effects of L1C products.

The different methods implemented in this class are specific to database management. They make it possible, for example, to retrieve and to provide information to users on the number of products available locally (*info*) or the occupied disk space (*size*). It is also possible to clean the corrupted products in the database (*clean_lib*), to delete specific L1C (*remove_l1c*) or L2A (*remove_l2a*) products, or to generate product quicklooks (*compute_ql*).

### 4.2 Class 'Job'

The Sen2Chain tool was designed to facilitate the management of a large database of images and indices. To achieve this, it was necessary to simplify and automate numerous recurring tasks, as individual products (single dates or tiles) could not be treated individually due to their large number. The 'Job' class allows creating, editing, saving, and launching routines, from downloading to producing indices on one or more tiles. A 'Job' is used to execute the main functions of Sen2Chain at once, and can be scheduled to run at regular time intervals (using the Linux standard cron scheduling system).

A configuration file is required to plan the different actions to be performed. It is divided into two parts: 1) a first one with global parameters, such as the number of cores to use, the initial cron setting, the number of attempts for downloads, the creation of a log file, or even the cleaning of the database before or after execution, and 2) a second part consisting of a dataframe of the tiles to be processed, the time period and maximum cloud cover, and the actions to be executed (i.e., downloads, production of L2A, cloud masks and indices). It also allows the removal of intermediate products to save disk space (L1C / L2A).

### 4.3 Class 'TimeSeries'

The 'TimeSeries' class is used to simplify the extraction of time series. The user specifies i. his sites of interest, defined as points or polygons in shapefile vector format, ii. a list of the radiometric indices available via Sen2Chain, and iii. the first and last dates delimiting his period of interest. In addition, the user can choose different cloud masks, with the flexibility to define its parameters according to the specific needs of the analysis. Temporal extractions are then made according to the user's criteria and preferences, while available computing resources are optimized through the use of parallelization. The output result is a simple dataframe with a synthesis of all indices value statistics included inside vectors and time frame requested, saved in text CSV format.

## 5. Sen2Extract: a Web interface for extracting time series of environmental indices

Sen2Chain has originally been developed to meet the needs of specific research projects. Although often used in this specific context, it has also been used for larger operational requirements with data processing launched routinely every day (Mouquet et al., 2020).

The Sen2Chain tool was initiated in 2017 as part of the 'Sentinel-2 for Malaria Surveillance (S2Malaria)' project funded by CNES. This project aimed to use S2 data to help monitor environment-related malaria in Asian and African countries, including Madagascar and South Africa at a local scale where other environmental and meteorological information is not timely accessible. During this project, the online tool Sen2Extract was developed to facilitate the extraction of time series of environmental indices by users with no remote sensing expertise, using a simple and user-friendly interface (Figure 3). This tool, coupled with Sen2Chain and an index database on remote servers, was also designed to be queried by users from Asian and African countries where networks are sometimes limited. The tool was developed in R, with an online interface based on the R 'Shiny' package (Chang et al., 2020). The code is open, enabling its re-implementation and adaptation to users according to their needs. One instance of Sen2Extract is currently running at SEAS-OI Station and is accessible freely. It can be used to extract all the images and products that have been processed with Sen2Chain, at SEAS-OI, mostly over the Southwestern part of the Indian Ocean.

Sen2Extract allows specifying a region of interest by uploading a vector file (a zipped shapefile with points and/or polygons), choosing one or multiple indices to extract using a checkbox, and selecting a time range with the calendars provided (Figure 3). Finally, the user provides an email address to receive the calculation results or instructions for download. The graphical panel allows the user to navigate on the map, and identify geographic areas (tiles) for which indices are available in the database. By clicking on any tile, information is given on the dates and numbers of products available, as well as a quicklook of the last L2A image generated (which informs on the quality regarding cloud coverage). Extractions with Sen2Extract can only be made for geographic areas and time periods where data is available, or the tool will return an empty query.

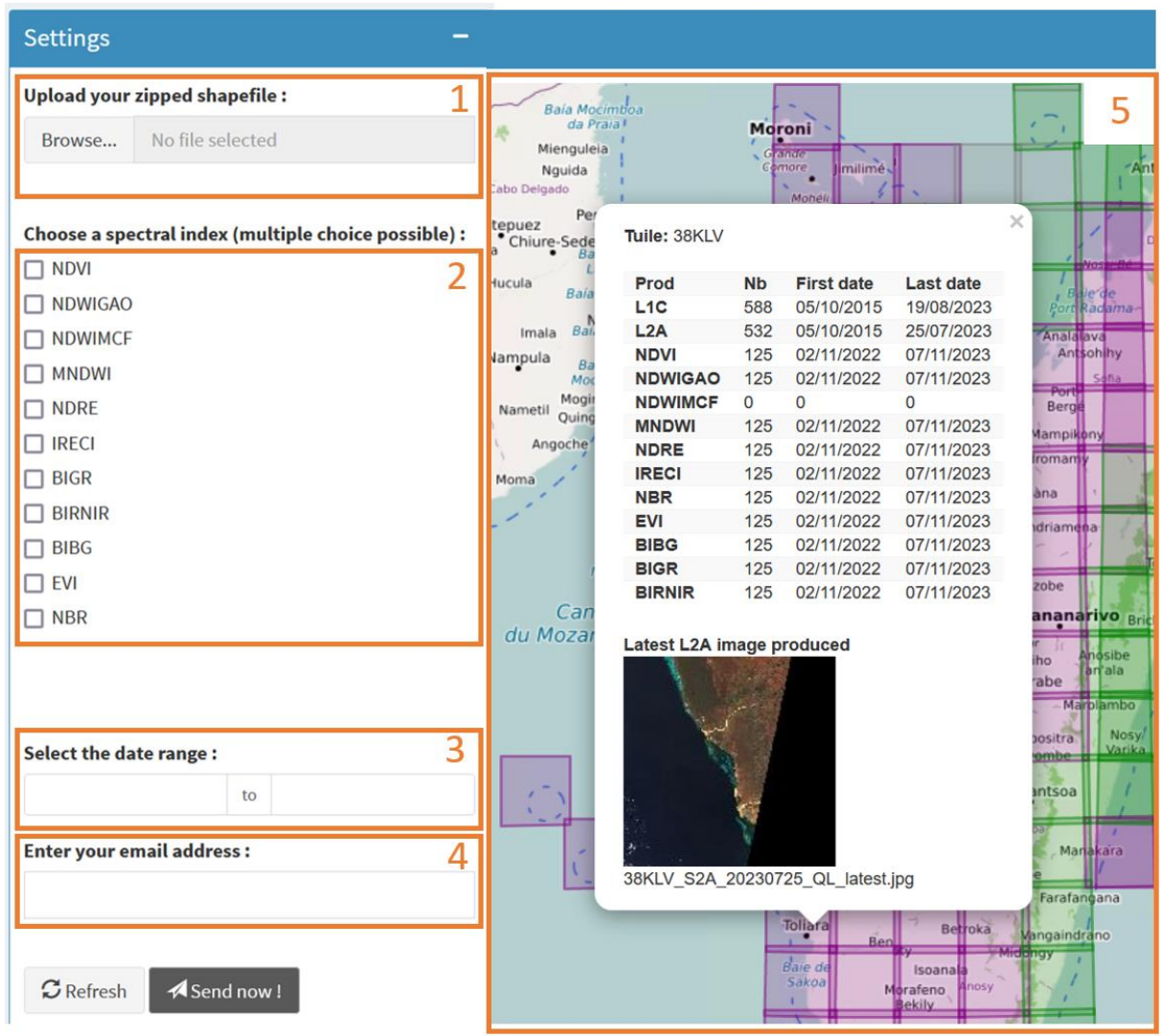

Figure 3: Steps for using Sen2Extract Web interface: 1 - uploading of the vector layer to extract indices; 2 - selection of the spectral indices; 3 - specification of the temporal range; and 4 - indication of the email address to receive instructions to download the processed data. The map panel (5) shows if images and products are available in the database for a specific tile (with the example of the S2 title 38KLV in the southwest of Madagascar).

The submission of a request via the Sen2Extract web interface triggers an extraction process on Sen2Chain and a query to the local database. Once completed, the user receives an e-mail containing an FTP download link to retrieve the results. The output data is provided in the form of a light .csv text file containing all the attribute characteristics of the input spatial entities, polygons or points, together with descriptive statistics for the selected indices (e.g. mean, minimum, maximum, standard deviation). The size of the output .csv file depends on the number of input vector spatial entities and the duration of the requested time period, with one line per entity and per date (Figure 4).

| | A | B | C | D | K |
|---|---|---|---|---|---|
| 1 | date | fid | tile | filename | mean |
| 83 | 14/12/2018 04:01 | 0 | 46QHD | S2A_MSIL2A_20181214T040141 | 2167.654278488127 |
| 84 | 19/12/2018 04:01 | 0 | 46QHD | S2B_MSIL2A_20181219T040149 | 1561.5949838415063 |
| 85 | 24/12/2018 04:01 | 0 | 46QHD | S2A_MSIL2A_20181224T040151 | 1777.1685401152172 |
| 86 | 29/12/2018 04:01 | 0 | 46QHD | S2B_MSIL2A_20181229T040149 | 1680.7137838977096 |
| 87 | 03/01/2019 04:01 | 0 | 46QHD | S2A_MSIL2A_20190103T040141 | 1648.0810032316988 |
| 88 | 13/01/2019 04:01 | 0 | 46QHD | S2A_MSIL2A_20190113T040121 | 1419.7672474357173 |

Figure 4: Extract of an output csv file with information about the geographic object used for extraction ('fid').

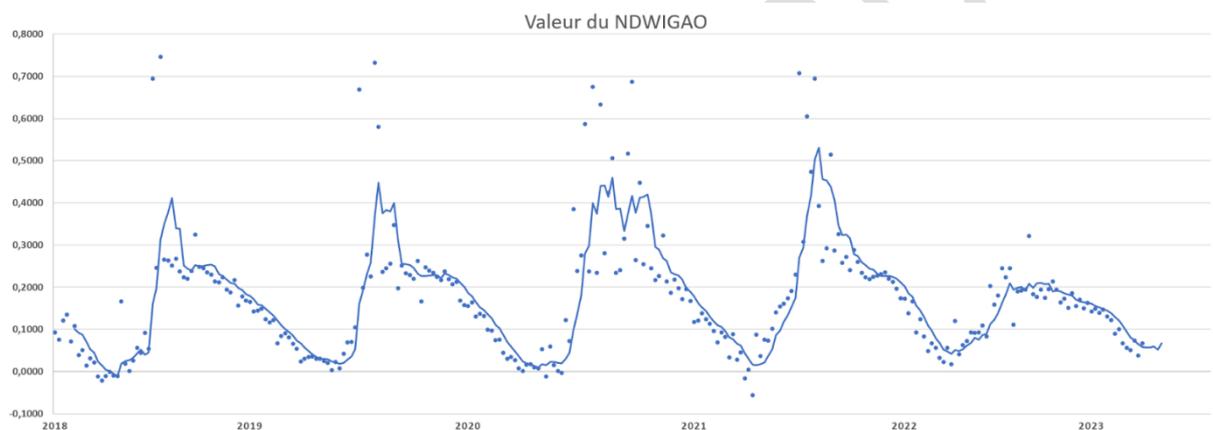

Figure 5: Visualisation of a time series of the NDWIGAO index extracted for a user-specified polygon in Myanmar. The x-axis represents time (in years) and the y-axis the NDWIGAO value. The trend line derived from the point values shows a clear seasonal variation.

The data extracted can be used for applications requiring time series analysis (Figure 5), such as for developing disease early warning systems, for the study of crop cycles, for the detection of environmental changes or for measuring the impacts after natural disasters (cyclonic floods for example). In this perspective, the LeptoYangon platform has been developed by the SCO Climhealth project to produce indicators of the suitable environment for leptospirosis (a bacterial environmental disease) transmission in the vicinity of Yangon city in Myanmar. This platform makes an operational use of S2 images through Sen2Chain and Sen2Extract to automatically produce risk maps every 5 days, as soon as a new S2 image is available in this area.

## 6. Discussion

The Sen2Chain tool was developed as a generic tool that can access and process large amounts of S2 imagery to support work in a wide variety of fields. To this end, the creation of the Job composite class allowed chaining and automating the processing of S2 images, making the tool more practical for regular use. Such a tool could for instance be used to feed observation networks or early warning systems.

Sen2Chain is original in that all developments have been carried out in Python language. Using Python and keeping the tool's codes open-source benefit from the presence of a very active developer community making available and regularly updating libraries for all kinds of applications. Advanced Python users can also freely use the tool's features to create their own production scripts and add new functionalities.

Sen2Chain offers a more complete set of functionalities than existing applications dedicated to the processing of S2 imagery, such as sen2r (Ranghetti et al., 2020). For instance, it integrates an advanced imagery download tool, supports an easy management of a database of images and derivatives, allows for automatic extractions of statistics in regions of interest (ROI), automates the processing in the form of tasks, and enables the processing of large geographical areas automatically.

However, Sen2Chain is not a versatile remote sensing image processing program like OTB or ENVI, and cannot be used as such for advanced computations. For the time being, Sen2Chain is dedicated to S2 data and the production/extraction of indices' times series from a local database that can contain images covering a large spatial-temporal domain.

Note that, as ESA/Copernicus production and storage capacities continuously increase, the L2A process level of Sentinel-2 data (Bottom-Of-Atmosphere images corrected for atmospheric attenuations), directly processed by Copernicus on its servers using Sen2Cor, started to be available online at a global scale. It is therefore envisaged in an upcoming Sen2Chain upgrade to integrate the possibility of directly retrieving those calibrated data instead of the original L1C level, thus saving resources and processing time.

Sen2Chain was initially developed to meet the needs of specific research projects and users. Its generic nature, and its potential to be applied in diverse contexts and over large geographical areas led to the opening and its code for greater dissemination and use. This tool constantly evolves depending on the functional needs, and updates providing improvements and new features are regularly released.

In the future, other tools for correcting atmospheric effects or detecting clouds could be tested. For example, the integration of NASA's (National Aeronautics and Space Administration) LaSRC algorithm, which uses a generic approach to derive surface reflectance over land from a variety of sensors and based on the 6SV radiative transfer model and the MODIS-derived red to blue band ratio (Vermote et al., 2016; Vermote et al., 2018), could improve the analysis of time series of indices with its better intercalibration of the images (Chraibi et al., 2022). The use of the Fmask algorithm (Qiu et al., 2019) could also be applied to the data to derive clouds and cloud shadows, with the aim of reducing outliers resulting from poor cloud detections in time series.

Many other smaller adjustments could improve the operation of the chain. These include a more explicit link with a database management system (DBMS) for better collection management and accelerated querying, better management of duplicate data (locally, or online), the development of local data cleaning tools to save disk space, such as being able to erase L1C, L2A, cloud masks, indices products on any or part of the temporal range and according to cloud cover criteria. The 11 spectral indices generated by Sen2Chain responded to specific users' needs but additional indices could easily be provided to meet the needs of other users/applications (see for instance the more extensive SentinelHub list of spectral indices that can be generated from S2 images).

Sen2Chain uses and is largely dependent on Python libraries and external tools (EODAG, Sen2Cor) for its operation. These tools are updated more or less frequently according to the needs and possibilities of programmers, to correct bugs, add functionalities, or for compatibility concerns (data, libraries, providers). Sen2Chain must therefore follow the same pattern of regular updates to continue to be operational.

## 7. Conclusions

The increasing volume of publicly accessible remote sensing data brings many opportunities, but also operational challenges. Developing processing chains that can automate the processing of large databases of remote sensing data, such as Sen2Chain, is increasingly important to fully benefit from the time series generated by today's and tomorrow's satellites. Sen2Chain makes it possible to handle time-consuming and repetitive processing of S2 images by offering parallelization and automation functions. If its operation is relatively simple, few tools offer the possibility of easily extracting large time series of spectral indices and for specific regions of interest. Developed primarily for scientific needs, the free and open-source nature of Sens2Chain enhances its accessibility and availability, and broadens its potential use.

The development of this open solution is part of long-term use perspectives as S2 imagery is expected to be available for many years to come, with the forthcoming launch of the new Sentinel-2 C and D satellites with similar specifications (in September 2024 for Sentinel-2 C). Moreover, the next generation of Landsat satellites should also produce images with characteristics similar to those of Sentinel-2, making high resolution images of any point on the globe available on a daily basis. This will further extend the uses of these images across a wide range of disciplines and applications (i.e., monitoring of phenomena that are changing rapidly).

## 7. Acknowledgments

Thanks to UMR Espace-Dev colleagues Jean-François Faure, Laurent Demagistri, Christophe Charon, and Thibault Catry for their help testing, deploying, and promoting the tool.

## Computer code availability

Sen2Chain was developed under the GNU license version 3 (GPLv3+). The tool is entirely coded in Python. The tool is freely available at this address: https://framagit.org/espace-dev/sen2chain. Installation and use of the tool are documented at this address: https://sen2chain.readthedocs.io/en/latest/ . Contact: tools@seas-oi.org

## Funding

This work was supported by:

- The S2Malaria project funded by CNES (Appel à Propositions de Recherche 2017, 2018 and 2019)
- The EU Interreg V project Renovrisk-Impact 14718 funded by the European Union, IRD, and the Reunion Regional Council.
- The Climhealth project (2020-2021) funded by CNES and accredited by the Space Climate Observatory France

List of Figures

List of Tables